\documentclass[prl,twocolumn,aps]{revtex4}
\usepackage{graphicx}
\usepackage{amsfonts}

\usepackage{amsfonts}
\usepackage{amssymb}
\usepackage{amsmath}
\usepackage{dsfont}
\usepackage{hyperref}
\usepackage{slashed}
\usepackage{empheq}
\usepackage{multirow}
\usepackage{fancyhdr}
\usepackage{appendix}
\usepackage{subfigure}
\usepackage{times}

\newcommand{\be}{\begin{equation}}
\newcommand{\ee}{\end{equation}}
\newcommand{\beq}{\begin{equation}}
\newcommand{\eeq}{\end{equation}}
\newcommand{\bea}{\begin{eqnarray}}
\newcommand{\eea}{\end{eqnarray}}

\def\pd{\partial}

\def\a{\alpha}

\def\p{\phi}
\def\te{\theta}

\DeclareMathOperator{\hz}{h\relax{\kern-.15em}z}
\DeclareMathOperator{\pz}{\psi\relax{\kern-.15em}z}

\begin{document}
%\preprint{MCTP-12-xx}

\title{ A String Theory Explanation for Quantum Chaos in the Hadronic Spectrum}

\author{Leopoldo A. Pando Zayas}\email{lpandoz@umich.edu}
\affiliation{Michigan Center for Theoretical Physics, University of Michigan, Ann Arbor, MI 48109, USA}

\author{Dori Reichmann}\email{dorir@umich.edu}
\affiliation{Michigan Center for Theoretical Physics, University of Michigan, Ann Arbor, MI 48109, USA}

\date{\today}
\begin{abstract}
In the 1950's Wigner and collaborators provided an explanation for the spectrum of hadronic excitations in terms of Random Matrix Theory.  In the 1980's  it was understood that some hadronic spectral properties were generic to systems whose classical limit is chaotic. We use string theory to demonstrate explicitly how, under very general conditions, recent holographic models of strong interactions have a spectrum compatible with Wigner's  conjecture.
\end{abstract}
\pacs{11.25.Tq,  05.45.-a,  11.30.Na}

\maketitle

\section{Introduction}

During the 1950s Eugene Wigner set out to describe the general properties of energy levels of highly excited states of heavy nuclei \cite{0862.01040}. The main idea was to describe such a complex nuclear system by a Hermitian Hamiltonian $H$ and to connect the results to Random Matrix Theory (RMT) \cite{Mehta:Book}. Wigner proposed regarding a specific $H$ as behaving like a large random matrix that is a member of an ensemble of Hamiltonians, all possessing similar general properties. Consequently, the spacings between energy
levels of heavy nuclei could be more easily modelled by the spacings between
successive eigenvalues of a random $N\times N$-matrix as $N\to \infty$.

During the 1980's a connection between RMT evolving from Wigner's work and quantum chaos was established through the understanding of simple and universal properties of the energy level fluctuations. Spectral fluctuations of quantum systems whose classical limit is fully chaotic show a strong level repulsion that depends only on the symmetries of the system, and agrees with the predictions of RMT \cite{Bohigas:1983er}.  In contrast, classically integrable systems give rise to uncorrelated adjacent energy levels, that are well described by Poisson statistics  \cite{1977}. Haq, Pandey and Bohigas (HPB) demonstrated that the fluctuation properties of nuclear energy
levels are in agreement with RMT \cite{PhysRevLett.48.1086}. HPB considered data consisting of $1407$ resonance energies corresponding to $30$ sequences
of $27$ different nuclei \cite{PhysRevLett.48.1086}; the connection with quantum chaos was then explored \cite{Bohigas:1989rq}. A statistical analysis looked at the experimentally measured mass spectrum of hadrons $(N,\Delta, \Lambda$ and $\Sigma$ and
all the mesons up to $f_2(2340))$ taken from the Particle Data Group Summary Tables and  concluded that the nearest-neighbor mass-spacing
distribution of the meson and baryon spectrum is described by the Wigner surmise corresponding to the Gaussian Othogonal Ensemble (GOE) \cite{Pascalutsa:2002kv}. Lattice studies of QCD and its supersymmetric versions  found similar results based on eigenvalues of the
Dirac operator \cite{Markum:2005ft,Bittner:2004ff}.

Here, by studying the spectrum of hadronic states in holographic models developed in string theory, we provide an explicit realization of Wigner's conjecture.

We first discuss the essential ingredients of the gauge/gravity duality and the mini-superspace approximation to the spectrum of certain hadronic states in the dual string theory. We then calculate the spectrum of eigenvalues and demonstrate that the level spacing is well described by the GOE;  we also comment on an observation regarding the spectrum that might be a smoking gun of holographic models. We provide a series of appendices with technical details of various calculations.

%%%%%%%%%%%%%%%%%%%%%%%%%%%%%%%%%%%%%%%%%%%%%%%%%%%%%%%%%%%
\section{ Quantum chaos in the holographic hadronic spectrum}\label{Sec:Hadrons}
%%%%%%%%%%%%%%%%%%%%%%%%%%%%%%%%%%%%%%%%%%%%%%%%%%%%%%%%%%%%%%%%%%%%%%%%%%%%%%%%%%%%%%%%
The gauge/gravity correspondence has provided a very special window into the nature of strongly coupled field theories by mathematically
identifying them with  dual string theories \cite{Maldacena:1997re,Witten:1998qj,Gubser:1998bc}.  The correspondence draws on a set of ideas dating back to 't Hooft who argued that, in the appropriate limit, a field theory can be described by strings \cite{'tHooft:1973jz}. Since its inception, more than a decade ago, an important goal has been the extension of the correspondence to the case of theories with properties similar to those displayed by Quantum-Chromodynamics (QCD) in the strongly coupled regime; hoping to explain properties that are experimentally verified but defy a theoretical explanation, like confinement and chiral symmetry breaking. The gauge/gravity correspondence has produced various models of confinement and highlighted the universal behavior that follows
from demanding that the dual Wilson loop satisfy the area law \cite{Brandhuber:1998er}. Remarkable
progress in matching to the spectrum of light hadrons in QCD has been achieved using one of these models \cite{Sakai:2004cn}.

An organizing principle of holographic dualities is that the energy scale in field theory is geometrized by an extra spatial dimension on the gravity side. Objects localized at large values of the  $r$-coordinate are identified with states localized in the ultraviolet of the field theory. Similarly, objects localized at small values of the $r$-coordinate are identified with states defined in the infrared of the field theory. Composite hadronic states that emerge from strong dynamics live mostly in the neighborhood of the smallest possible values of the $r$-coordinate, $r_{min}$, which are naturally identified with the strong coupling scale $\Lambda_{QCD}$ in the field theory.

\begin{figure}[h!]
\begin{center}
\resizebox{2.5in}{!}{\includegraphics{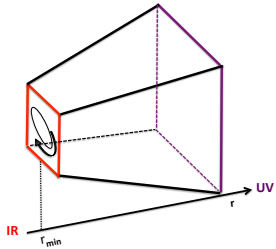}} \,\,\,\,
\caption{\label{Fig.Holography} The holographic principle associates the direction $r$ in the bulk with the energy scale in the field theory.}
\end{center}
\end{figure}

In the gauge/gravity paradigm a classical trajectory in string theory serves as a description of a quantum state in the dual field theory. The oldest realization
of this paradigm is given by Regge trajectories which precedes the AdS/CFT by more than thirty years. The old argument explaining Regge trajectories as spinning strings is now modified in the gauge/gravity correspondence (see Fig. (\ref{Fig.Holography})). The classical trajectory whose conserved quantities (angular momentum $J$
and energy $E$) satisfy $J\sim E^2$ is interpreted as describing the principal Regge trajectory, that is, a
sector of the spectrum of hadrons with the same relation where now $J$ is the spin and $E=M$ is the mass of
the hadron. Crucially, the string spins deep into the bulk geometry. The gravitational red shift between the small  $r$ region and the boundary ($r\to \infty$) converts the string scale $\alpha'$ to the QCD scale. More precisely, the QCD string tension is given in terms of the geometry as $T_{QCD}\sim g_{00}(r_{min})/(2\pi \alpha')$ highlighting the role of the holographic direction.

One fact that has largely dictated what can be done in the context of string theory is the fact that it is, at the moment, a technically unsurmountable problem to solve for the spectrum of string theory in Ramond-Ramond backgrounds. This situation has lead to an emphasis on semi-classical quantization. In the last ten years, since the insightful work of Berenstein-Maldacena-Nastase \cite{Berenstein:2002jq} and also Gubser-Klebanov-Polyakov \cite{Gubser:2002tv}, the semiclassical analysis of the spectrum has dominated the field. Further  advances in the case of ${\cal N}=4$ CFT and its interpretation from the dual $AdS_5\times S^5$ have precisely relied in this semiclassical approach to the spectrum \cite{Beisert:2010jr}.  Unfortunately, such semi-classical quantization has largely limited itself to integrable solutions even in the case of non-conformal backgrounds \cite{Gimon:2002nr,Bigazzi:2004ze}.

%%%%%%%%%%%%%%%%%%%%%%%%%%%%%%%%%%%%%%%%%%%%%%
\subsection{Quantum spectrum from minisuperspace in string theory}
%%%%%%%%%%%%%%%%%%%%%%%%%%%%%%%%%%%%%%%%%%%%%%%%%
In standard quantum mechanics, starting with a classical picture there is a definite prescription for finding the energies of a system.  We start with a classical Hamiltonian $H$, promote the generalized momenta and coordinates to operators and solve the Schr\"odinger equation.  The eigenvalues are the allowed energies of the system.   In our new, holographic picture, we are faced with new classical data -- the trajectories of strings.  How are these trajectories  to be mapped to a quantum problem, and how do we extract information on the resulting spectrum?  In string theory the Virasoro constraint provides the analog to the Schr\"odinger equation.  It leads to the mass shell condition and is usually written as $\left({\cal H}=(L_0-a)\right)|\Psi>=0$, where $L_0$ is a Virasoro generator and $a$ is a constant resulting from normal ordering.  In this framework we think of $|\Psi>$ as a the precise analog of the wave function.

We will follow a simplification known as the mini-superspace, an approximation where only modes corresponding to the center of mass are retained. The minisuperspace idea was originally introduced in the context of quantum cosmology \cite{Hartle:1983ai}; in string theory, or more precisely, Liouville theory it was formulated in  \cite{Seiberg:1990eb}. More recently the mini-superspace has played a clarifying role in more complicated setups \cite{Douglas:2003up}.

We start with the Polyakov action in the conformal gauge:
\be
\label{Eq:String}
S=-\frac{1}{2\pi \alpha'} \int d\tau \int d\sigma \, G_{MN}\left(-\dot{X}^M\dot{X}^N+X'{}^MX'{}^N\right),
\ee
where $M=0, \ldots, 9$ and $1/2\pi \alpha'$ is the fundamental string tension. To implement the minisuperspace approximation, we consider an Ansatz where only one function depends on the world sheet coordinate  $\sigma$, let us call it $X^9$ and the rest depend on time, $\tau$:
$X^M=\left(x^n(\tau),X^9(\sigma)\right)$. After integrating equation (\ref{Eq:String}) with respect to $\sigma$ we obtain essentially a particle-like Hamiltonian of the form:
\bea
\label{Eq:metric-mini-superspace}
{\cal H}&=&\frac{1}{2}g^{mn}p_m p_n+ \frac{1}{2}V(x), \quad
g_{mn}(x)= \int\limits_0^L  d\sigma G_{MN}(x), \nonumber \\
V(x)&=&\int\limits_0^L  d\sigma G_{99}(x)X'{}^9 X'{}^9,
\eea
where $p_m$ are the canonical momenta conjugate to the coordinate $x_m$ and we have assumed that the metric $G_{MN}$ is diagonal.

The quantum spectrum is determined, in the framework of the minisuperspace formalism, by the following equation:
\be
\label{Eq:TheProblem}
-\Delta \Psi + V(x)\Psi=0,
\ee
where the Laplacian is computed in the metric $g_{mn}(x)$ given by equation (\ref{Eq:metric-mini-superspace}).  This approach accesses only a special kind of string states (quantizing the center of mass motion).  The minisuperspace  also leaves out other important sectors of superstring theory like those coming from fermionic fields or fluxes in the geometry rendering the results somewhat limited. Nevertheless, in some specific situations with a large amount of symmetry the full answer of the string spectrum can be obtained from the minisuperspace approximation \cite{Teschner:1997fv,Maldacena:2000hw}.

%%%%%%%%%%%%%%%%%%%%%%%%%%%%%%%%%%%%%%%%%%%%%%%%%%%%%%%%%%%%%%%
\subsection{Energy eigenvalues}
%%%%%%%%%%%%%

In what follows we will solve (\ref{Eq:TheProblem}) for two supergravity backgrounds dual to confining field theories.

The main idea of the holographic dictionary is to replace the study of a 4d field theory, in the appropriate limit, by that of a 10d gravity theory \cite{Maldacena:1997re}. In this paper we are interested in 10d metrics of the form:
\be
\label{Eq:10d}
ds^2=A^2(r)(-dt^2 +dR^2+R^2d\varphi^2+dz^2)+B^2(r)d r^2 +ds_5^2,
\ee
where $r$ is the so called holographic direction, $ds^2_5$ represents a compact 5d space and can depend on $r$; its integrated volume form is denoted by $\omega_5(r)$. The space where the field theory lives $(\mathbb{R}^{1,3})$ has been parametrized in cylindrical coordinates $(t,R,\varphi,z)$. The precise backgrounds we consider here are due to Maldacena-N\'u\~nez (MN)\cite{Maldacena:2000yy} and Witten (WQCD)\cite{Witten:1998zw}.

Rather than a closed spinning string at the minimum of $r$ corresponding to the holographic Regge trajectory (depicted in Fig. \ref{Fig.Holography}),  we will consider a winding string which is largely localized in the same region, $r_{min}$. Closed strings are generically mapped to glueball states; our spectrum will correspond, consequently, to the spectrum of glueballs. We hope that this spectrum forms the dominant part of the spectrum of hadrons but we are not able to directly apply our results to all hadrons.

\begin{figure}[h!]
\begin{center}
\resizebox{1.6in}{!}{\includegraphics{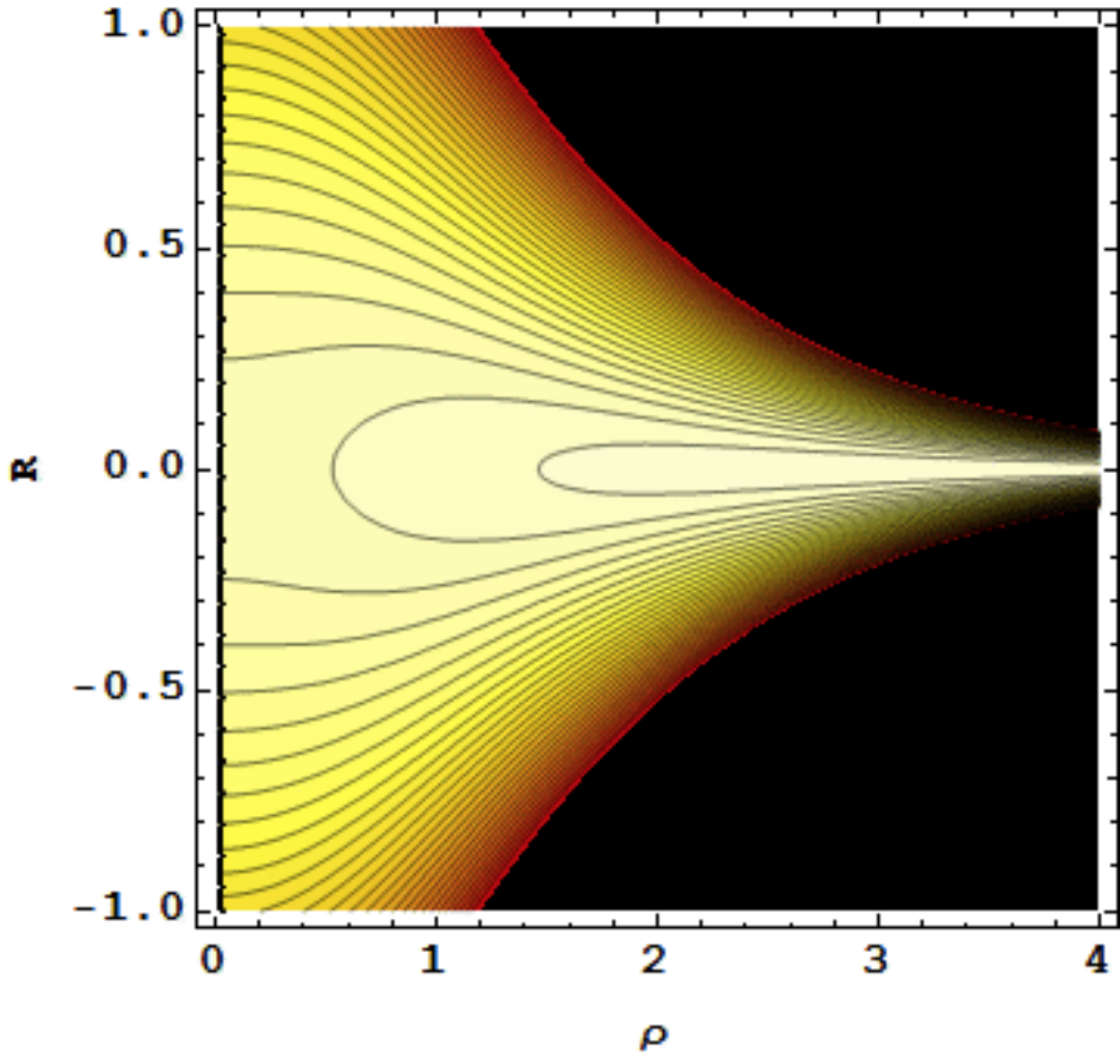}} \,\,\,\,
\resizebox{1.5in}{!}{\includegraphics{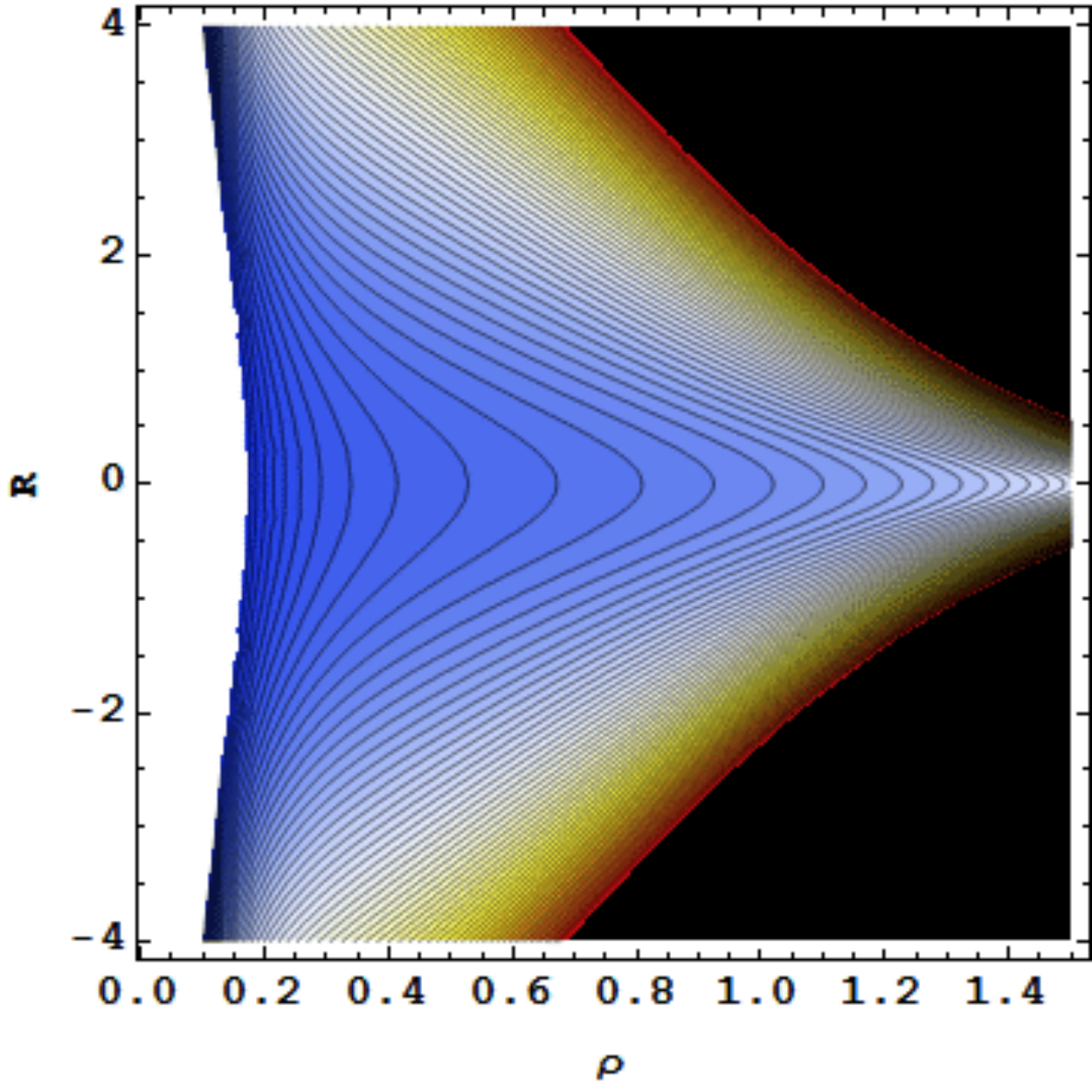}}
\caption{\label{FigPotentials} Density plots of the effective potential, $V(\rho, R)$, in the MN and WQCD backgrounds.}
\end{center}
\end{figure}

We focus on states for which $\varphi(\sigma)=\alpha\, \sigma$, that is, where the string winds $\alpha$ times around the $\varphi$ direction. {\it A priori} $\alpha$ is not a good quantum number in the backgrounds we consider as the string can unwind; we return to this issue later. In the formalism described in the previous section we set $X^9(\sigma)=\varphi(\sigma)$. After some simplifications (see appendix \ref{App:MSP}) we obtain a general problem of the form

\begin{equation}
\label{Eq:Spectral}
\left(-\partial_R^2-l(r)\partial_r\left(m(r)\partial_r\right)+\omega^2 R^2 A(r)^4\right)\Psi=E^2\Psi,
\end{equation}
where $l(r)=1/(A(r)B(r)\omega_5(r))$, $m(r)=A(r)^3\omega_5(r)/B(r)$ and $\omega=\alpha/(\pi \alpha')$. The functions $A(r), B(r)$ and $\omega_5(r)$ enter in the 10-d metric Eq. (\ref{Eq:10d}). To develop intuition into this spectral problem we comment on some of the most salient features at the heart of the results.
The explicit form of the effective potential, $V(\rho, R)$, in coordinates $(\rho(r), R)$ where the kinetic terms are canonical, can be found in appendix \ref{App:MSP}.  For both theories analyzed, the potential $V(\rho, R)$ is impenetrable as  $\rho$ approaches its spatial boundary. Near $\rho=0$ the MN potential has a finite piece which is independent of $R$ while the WQCD potential blows up as $-1/\rho^2$. The point that we would like to emphasize graphically is that the potentials provide a rough realization of a Bunimovich stadium \cite{springerlink:10.1007/BF01197884} (see figure \ref{FigPotentials}). In the figures we exploit the symmetries of the problem around $\rho=0$ where in the full geometry a closing cycle leads to a smooth origin of the form $\mathbb{R}^{3}$ for MN and $\mathbb{R}^2$ for WQCD. With this symmetry configuration in mind we should characterize both potentials as Bunimovich-like stadia with a bump in the center, $\rho=0$, \'a la Sinai. The similarity of our potentials with the typical potentials of quantum chaos (Bunimovich and Sinai) makes the appearance of quantum chaos more plausible.

%%%%%%%%%%%%%%%%%%%%%%%%%%%%%%%%%%%%%%%%%%%%%%%%%%%%%%%%%%%%%%%
\section{The spectrum of excited holographic glueballs}\label{Sec:Results}
%%%%%%%%%%%%%%%%%%%%%%%%%%%%%%%%%%%%%%%%%%%%%%%%%%%%%%%%%%%%%%%%%%%%%%

Quantum chaos studies the quantum properties of classically chaotic systems. In such systems the local statistics of the energy spectrum play a key role. An important feature is the level spacing distribution $P(s)$, which is the distribution function of nearest-neighbor spacings $E_{n+1}-E_n$ as we run over all levels plays a key role. A dramatic insight of quantum chaos is given by the universality conjecture for $P(s)$. If the classical dynamics is chaotic, then $P(s)$ coincides with the corresponding quantity for the eigenvalues of a suitable ensemble of random matrices $P(s)=s \exp(-s^2/M^2)$ \cite{Bohigas:1983er}.

We solve the mini-superspace spectrum  (\ref{Eq:Spectral}) for two prominent supergravity backgrounds dual to confining theories: MN and WQCD (see appendix \ref{App:Backgrounds} for details of the backgrounds.). The eigenvalue problem itself is solved numerically using spectral decomposition \cite{Boyd01chebyshevand}. We relegate most technical details to the appendices \ref{App:MSP} but note a few key features of the problem.
In the spectral method, the left hand side of Eq. (\ref{Eq:Spectral}) is written as a 2-dim matrix folding the two spatial dimension $R$ and $\rho$ together using Kronecker outer product. With a slight abuse of notations we call that matrix the Hamiltonian. The energies are the square roots of the Hamiltonian eigenvalues. We exploit the parity symmetry in the $R$ direction to sort the states by parity and conserve computational resources. Our basis for spectral decomposition in the $R$-direction uses Hermite functions, since both $\pd_R^2$ and $R^2$ appear as tridiagonal matrices. The complete Hamiltonian is non-zero only for elements that are within the $N_r+1$ sub-diagonal (where $N_r$ is the number of spectra functions used in the $r$ direction). The block-diagonal form of the Hamiltonian is very different from the dictum of Wigner-Dyson assumptions as summarized by RMT. We will track the implications of this difference.

After using the reparametrization freedom in $R$ and $\rho$ in both problems we are left with a single parameter $\alpha/(\pi\alpha')$ which measures the interaction strength, we arbitrarily set $\alpha/(\pi\alpha')=1$. This paremeter sets the energy scale at which the interaction term becomes important. Below that scale the Hamiltonian is, to a good approximation, a direct product and the eigenvalues will follow an uncorrelated Poisson distribution.

\begin{figure}[h!]
\begin{center}
\resizebox{1.5in}{!}{\includegraphics{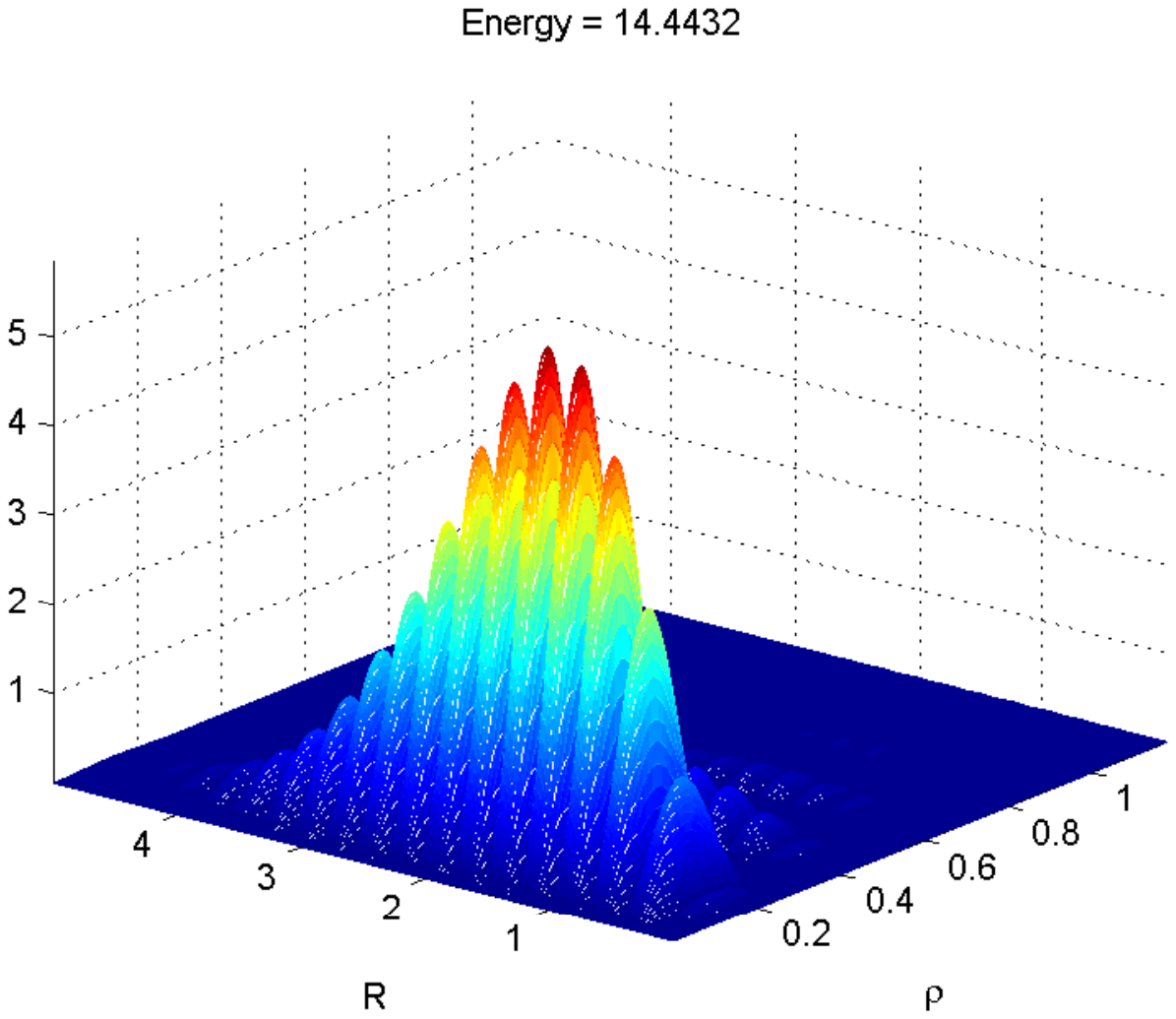}} \,\,\,\,
\resizebox{1.5in}{!}{\includegraphics{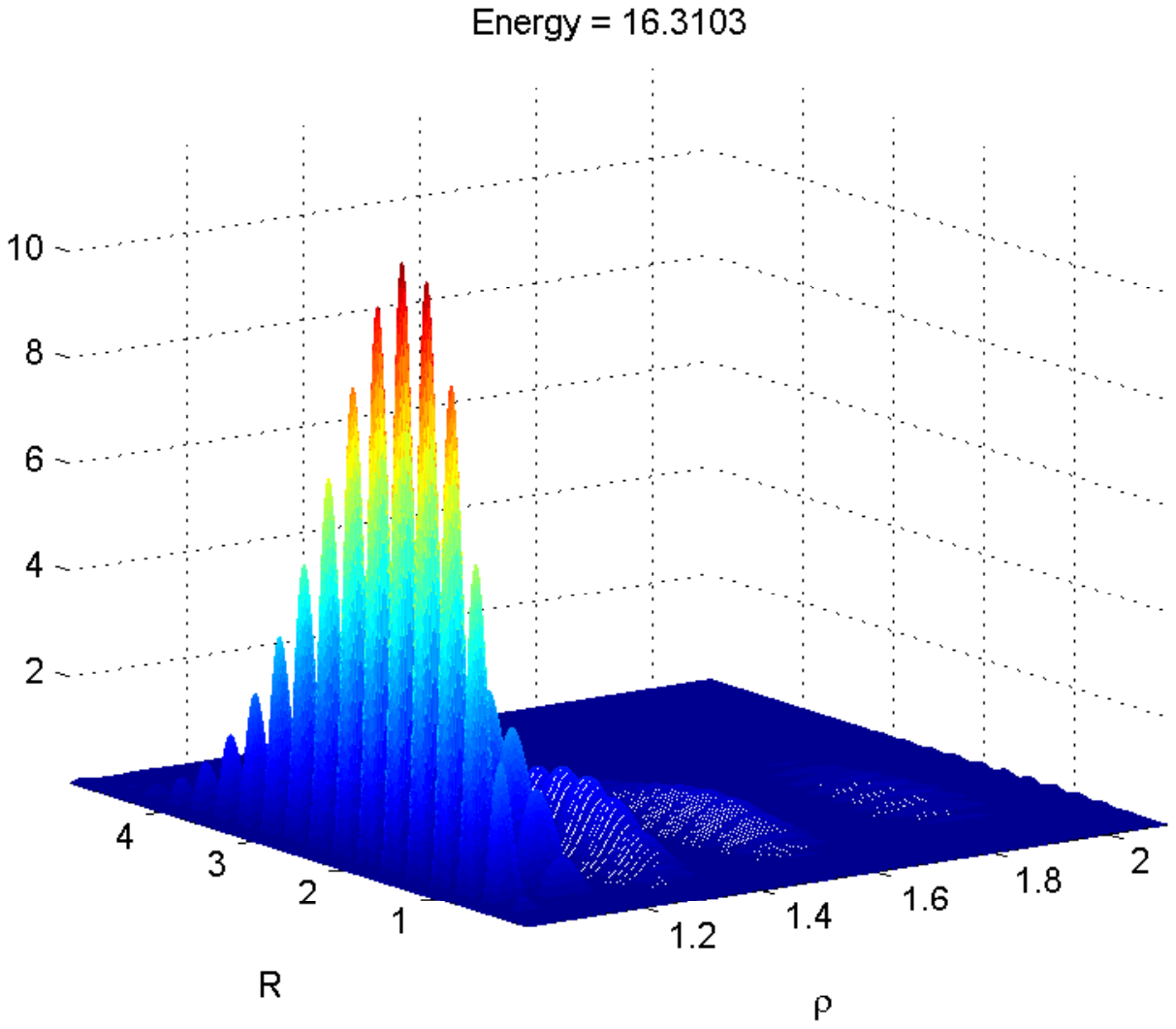}}
\caption{\label{Fig:wavefunctions}Wave functions. Smallness near $R=0$ guarantees that the mixing among states of different winding number is negligible. }
\end{center}
\end{figure}

In figure (\ref{Fig:wavefunctions}) we plot the absolute value square of a typical wave function for both backgrounds. The fast decay in $\rho$ is the direct consequence of the form of the potential in that direction (compared to the quadratic behavior in the $R$ direction). In both cases we can see that the value of the wave function near $R=0$ is small, this is not an accidental property of the Hermite functions. Rather, this indicates that the interaction term excites a large number of modes in the $R$ directions. This observation is reassuring, the strings we consider carry a fixed winding number, denoted by $\alpha$ before, around the origin. This winding number is not conserved near the tip, $R=0$, where the strings can unwind. Since the winding number is a good quantum number for strings that are far away from the tip, we expect the mixing of different winding modes to be proportional to the density of the wave functions near the tip which is shown in fig. (\ref{Fig:wavefunctions}) to be small.  Therefore the wave function smallness near $R=0$ suggests that the mixing of different winding sectors will be a small effect.

\begin{figure}[h!]
\begin{center}
\resizebox{1.5in}{!}{\includegraphics{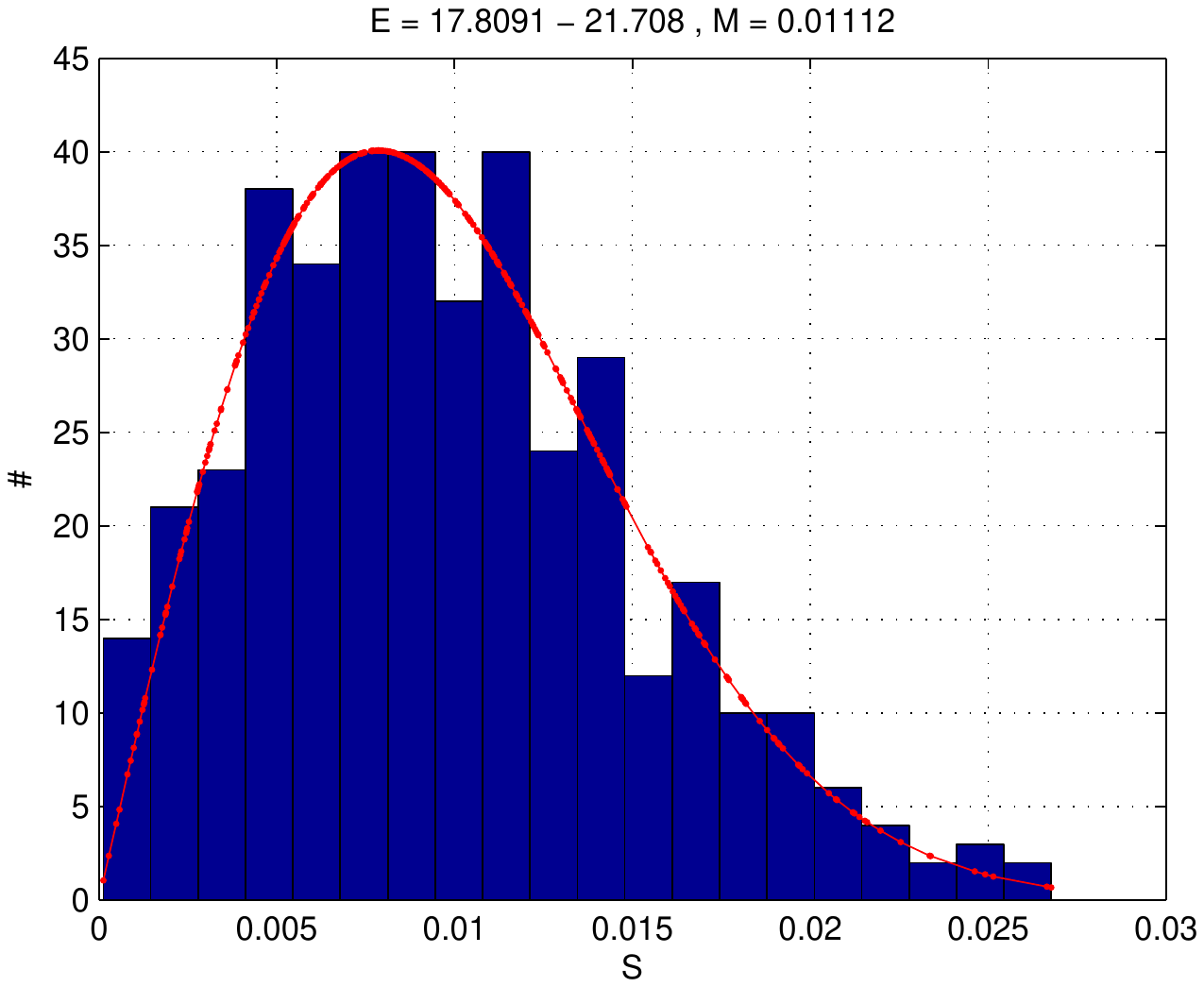}}
\resizebox{1.5in}{!}{\includegraphics{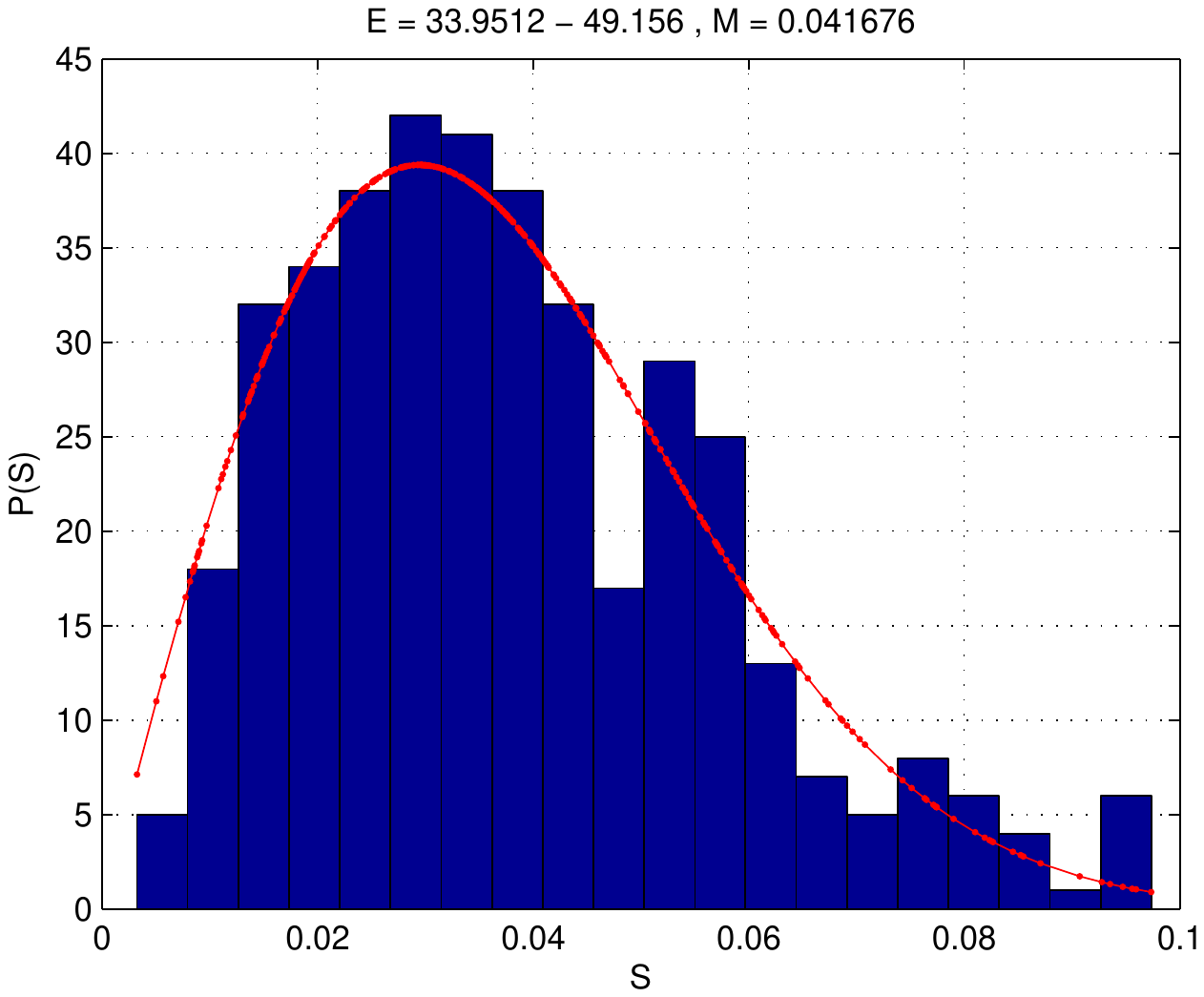}}
\caption{\label{Fig:Distribution}Eigenvalue spacing distribution $P(s)$ for the MN background and for the WQCD background.}
\end{center}
\end{figure}

After calculating the spectrum for both cases we are now ready to study the eigenvalue spacing distributions. First we choose an arbitrary energy range (above the range of small interactions) spanning about 400 energy levels. We calculate the energy difference between eigenvalues in the chosen range and plot them on an histograms, and compare them to Wigner distributions $P(s)\sim s \exp(-(s/M)^2)$. The results are plotted in Fig. (\ref{Fig:Distribution}). The  root mean square (RMS)  between Wigner's distribution and data is below $10^{-3}$ when the distribution is normalized so the sum is one. This excellent matching  to Wigner distribution proves our main claim that the spectrum of hadrons in the MN and WQCD theories shows a quantum chaotic eigenvalue distribution.

It is worth noting that the lowest lying eigenvalues lead to a Poisson distribution and we drop them from consideration. This is not the case in QCD but the discrepancy can be attributed to the fact that holographic models require a parametrically large $\Lambda_{QCD}$.

\begin{figure}[h!]
\begin{center}
\resizebox{1.5in}{!}{\includegraphics{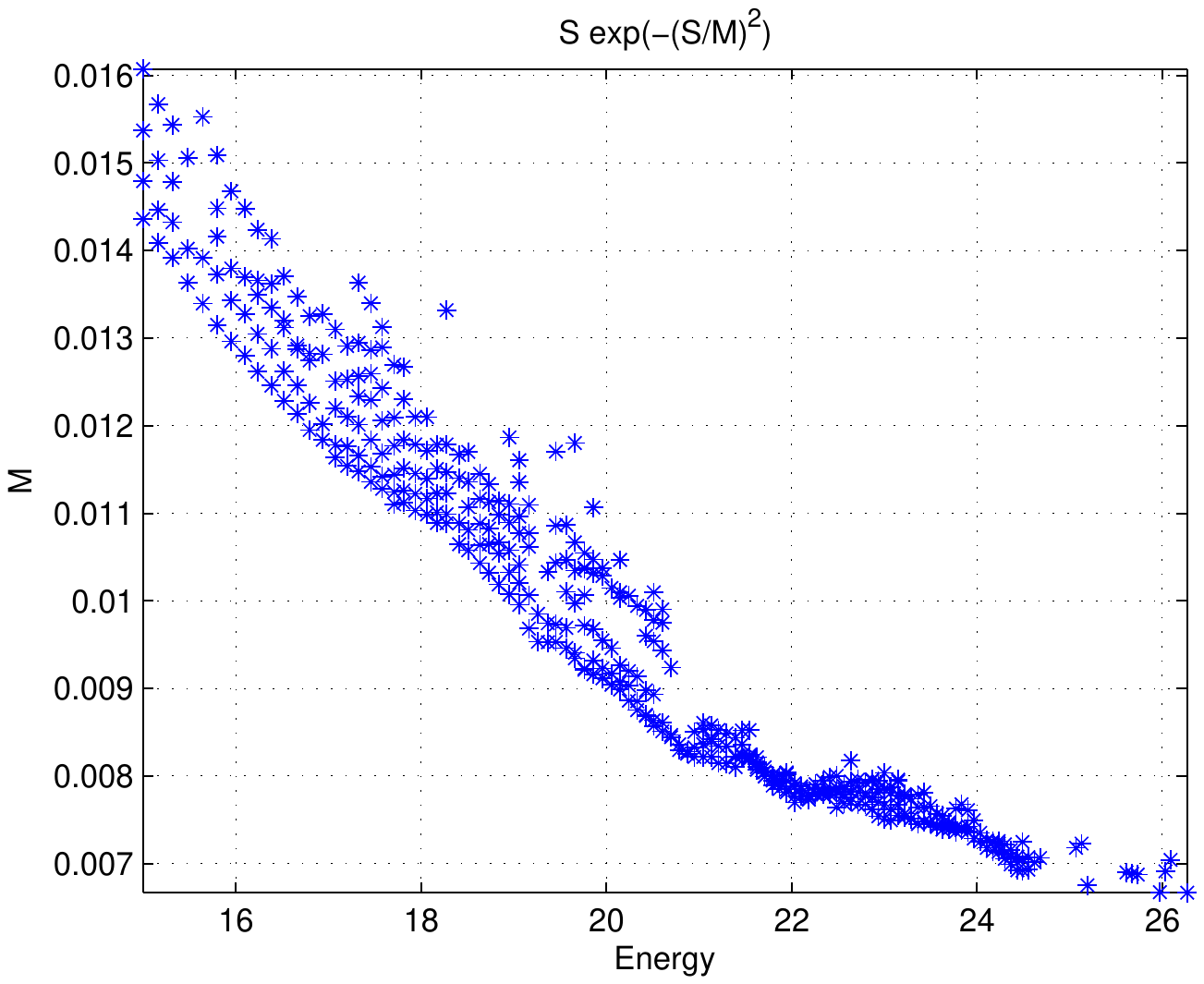}}
\resizebox{1.5in}{!}{\includegraphics{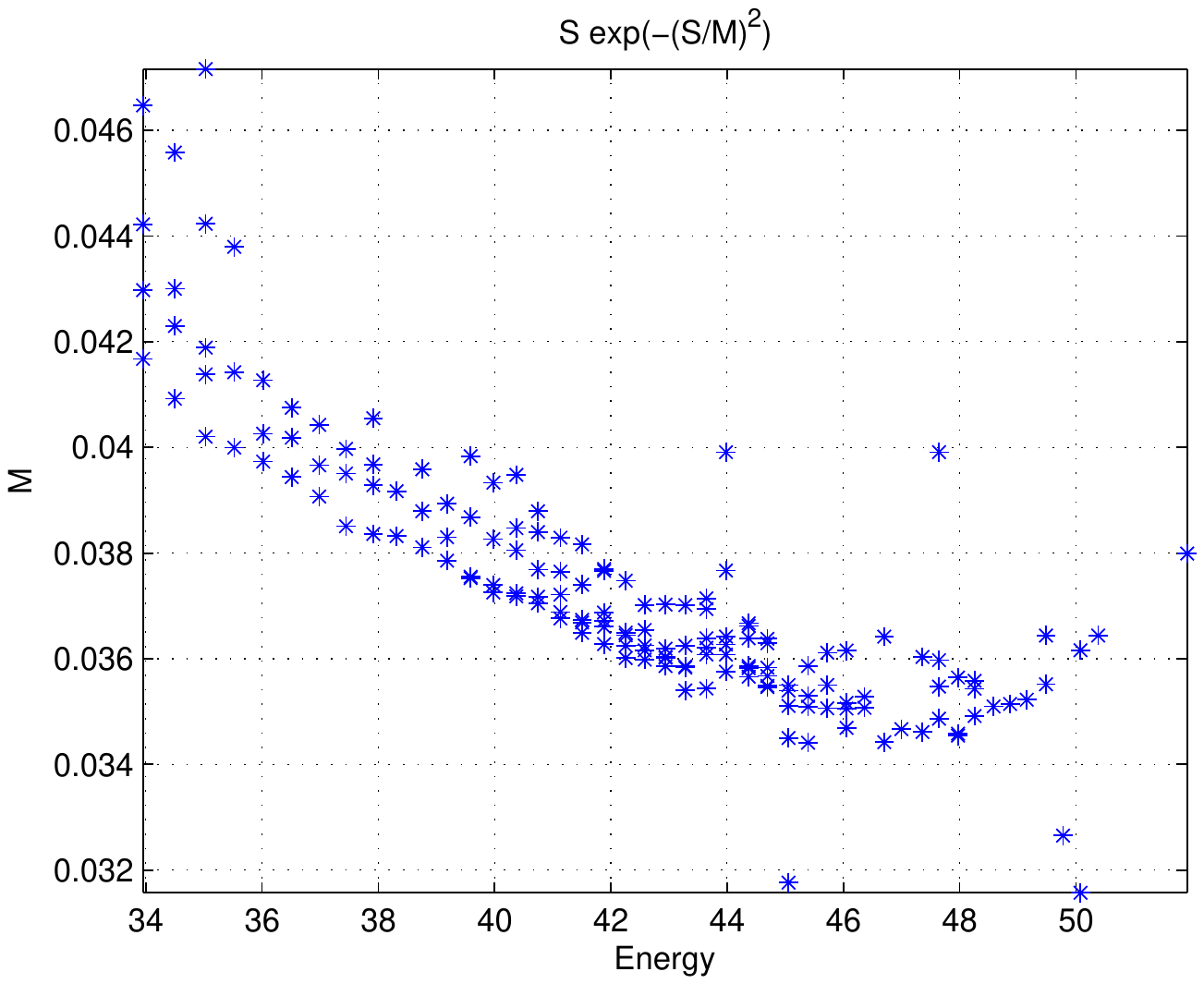}}
\caption{\label{Fig:MvE}Dependence of $M$ on the energy region for the MN background and for the WQCD background.}
\end{center}
\end{figure}

We repeat the above procedure for the entire energy spectrum we can accurately calculate. Varying energy ranges and bin sizes we determine the value of the parameter $M$ in the Wigner distribution as a function of energy (the mean energy of the energy range used for the histogram). We keep only the cases where the matching to the Wigner distribution was good (measured by RMS $<10^{-3}$). The results are displayed in Fig.  (\ref{Fig:MvE}). The main feature shown is $M$ decreasing with energy. This property differentiates between the models we study and the random matrices discussed in the Wigner-Dyson approach (where $M$ does not depend on energy). We attribute this result to the clustered block-diagonal nature of the Hamiltonians, however we cannot prove this connection. We conjecture that the dependence on energy $M(E)$ is universal to all string theory models for confining theories beyond the two we study.

%%%%%%%%%%%%%%%%%%%%%%%%%%%%%%%%%%%%%%%%%%%%%%%%%%%%%%%%%%%%%%%
\section{Conclusions}\label{Sec:Conclusions}
%%%%%%%%%%%%%

As in previous studies, the holographic direction, $r$,  has been crucial. Simple models of strings in flat space such as the original string approach of the 1960's would not lead to the results presented here. The  condition that geometrically the supergravity backgrounds have to satisfy two properties
which are sufficient for an area law for the dual Wilson loop has been our only input. Most of our results follow from the ``end of the space wall'', that is, the fact that $g_{00}(r_{min})$ has a non-vanishing minimum at the smallest value of the holographic direction $r$.  This behavior implies that for supergravity backgrounds like $AdS_5\times S^5$ dual to conformal theories, the spectrum will be completely different. As shown in \cite{Basu:2012ae}, the holographic area law condition alone implies chaotic behavior in the classical equations of motion and we have now explicitly verified that the quantum spectrum is chaotic in the standard sense of the GOE distribution.

We have shown that in the minisuperspace approximation, where only a few degrees of freedom are retained in addition to the center of mass motion, the system retains sufficient generality enabling it to capture experimentally verified physics. More generally our results motivate studies of properties of the spectrum that are not dictated by integrability.

We have discussed two of the most studied models of holographic confining theories, the MN and
WQCD models. We have found that the spectrum of nearest-neighbor eigenvalues has a Gaussian distribution consistent with the GOE. Interestingly, this is the distribution observed in nuclei, hadrons and realistic lattice studies of QCD and its supersymmetric versions.

More speculatively, we have pointed out to a property that might be specific to holographic models of confinement:  the dependence of $M$ on the energy range $E$, in the statistics of eigenvalue levels.

We have thus, using the principles of holography,  provided a concrete Hamiltonian for hadronic excitations with the properties conjectured by Wigner more than half a century ago. In a a holographic sense we have provided a concrete realization of Wigner's ideas. This is thus an explanation of quantum chaos or RMT in a sector of the spectrum of the hadrons in holographic models.

%%%%%%%%%%%%%%%%%%%%%%%%%%%%%%%%%%%%%%%%%%%%%%%%%%%%%%%%%%%%%%%%%%%5
\section*{Acknowledgments}
We thank C. Keeler, A. Pierce and J. Sonnenschein for comments and  C. N\'u\~nez  for ongoing collaboration. L.A.P.Z. is thankful to P. Basu, D. Das and A. Ghosh for collaborations on related topics and to the KITP and Aspen Center for Physics
for hospitality during various stages of this work.  L.A.P.Z. is also thankful to Mark Srednicki for encouragement. This
research was supported in part by the National Science Foundation under Grant No. NSF PHY11-25915 (KITP),  grant No. 1066293 (Aspen) and by Department
of Energy under grant DE-FG02-95ER40899 to the University of Michigan.

\appendix
%%%%%%%%%%%%%%%%%%%%%%%%%%%%%%%%%%%%%%%%%%%%%%%%%%%%%%%%%%%%%%%%%%%%%%%%%55
\section{\label{App:Backgrounds} Supergravity backgrounds}
%%%%%%%%%%%%%%%%%%%%%%%%%%%%%%%%%%%%%%%%%%%%%%%%%%%%%%%%%%%%%%%
\subsection {\label{App:MN}The Maldacena-N\`u\~nez background}
%%%%%%%%%%%%%%%%%%%%%%%%%%%%%%%%%%%%%%%%%%%%%%%%%%%%%%%%%%%%%%%

The MN background whose IR regime is associated with ${\cal N}=1$
SYM theory is that of a large number of D5 branes wrapping an
$S^2$ \cite{Maldacena:2000yy} (see also \cite{Chamseddine:1997nm}). To be more precise: (i) the dual field theory to this
SUGRA background is the ${\cal N}=1$ SYM contaminated with Kaluza-Klein (KK)
modes which cannot be de--coupled from the IR dynamics, (ii) the
IR regime is described by the SUGRA in the vicinity of the origin
where the $S^2$ shrinks to zero size.   The full MN SUGRA background
includes the metric, the  dilaton and the
RR three-form.   It can also be interpreted as an uplifting to ten
dimensions a solution of seven dimensional gauged supergravity
\cite{Chamseddine:1997nm}.  The metric and dilaton of the background are
\begin{align}\label{dfmn}
    ds^2=&~e^\p\left[\eta_{\mu\nu}dx^\mu dx^\nu+\a'g_s N \left(dr^2+ ds^2_5\right)\right],
\cr
    ds_5^2=&~ e^{2g(r)}(e_1^2+e_2^2)+{1\over 4} (e_3^2+e_4^2+e_5^2)),
\cr
    e^{2\p}=&~e^{-2\p_0}{\sinh 2r\over 2e^{g(r)}},
\cr
    e^{2g(r)}=&~r\coth 2r -{r^2\over \sinh^2 \, 2r}-{1\over
  4},
\end{align}
where $\mu,\nu=0,1,2,3$ and
\begin{align}
    e_1=&d\theta_1, \qquad e_2=\sin\te_1 d\p_1,
    \cr
    e_3=&\cos\psi\, d\te_2+\sin\psi\sin\te_2\, d\p_2 -a(r) d\te_1,
    \cr
    e_4=&-\sin\psi\, d\te_2+\cos\psi\sin\te_2\, d\p_2 -a(r) \sin \te_1d\p_1,
    \cr
    e_5=&d\psi +\cos\te_2\, d\p_2 -\cos\te_1d\p_1, \quad
    a(r)={r^2\over \sinh^2r},
\end{align}
where $\mu=0,1,2,3$, we set the integration constant  $e^{\phi_{D_0}}= \sqrt{g_s N}$   The 3-form is
\begin{align}
    H^{RR} =~& g_sN \bigg[ - {1\over 4} (w^1 -A^1)\wedge (w^2 - A^2) \wedge ( w^3-A^3)
    +\cr& + { 1 \over 4}
    \sum_a F^a \wedge (w^a -A^a) \bigg]
    \cr
    \label{a}
    A =~& { 1 \over 2} \left[ \sigma^1 a(r) d \theta_1
    + \sigma^2 a(r) \sin\theta_1 d\phi_1 +
    \sigma^3 \cos\theta_1 d \phi_1 \right]
\end{align}
and the one-forms $w^a$ are given by:
\be
w^1 + i w^2   = e^{ - i \psi } ( d \theta_2 + i \sin \theta_2 d \phi_2)  ~,~~~~~~~~~~
w^3 = d \psi +\cos \theta_2  d\phi_2
\ee
Note that we use notation where $x^0,x^i$ have dimension of
length whereas $\rho$ and the angles
$\theta_1,\phi_1,\theta_2,\phi_2,\psi$ are dimensionless and hence the
appearance of the $\alpha'$ in front of the transverse part of
the metric.

%%%%%%%%%%%%%%%%%%%%%%%%%%%%%%%%%%%%%%%%%%%%%%%%%%%%%%%%%%%%%%%%%%
\subsection{\label{App:WQCD}The Witten QCD background}
%%%%%%%%%%%%%%%%%%%%%%%%%%%%%%%%%%%%%%%%%%%%%%%%%%%%%%%%%%%%%%%%%%%

In this section we present the Witten QCD background; we include some notes on the holographic relations with the field
theory \cite{Witten:1998zw}.
The ten-dimensional string frame metric and
dilaton of the Witten model are given by

\begin{align}
\label{defns}
    ds^2=~&\left({r\over
    L}\right)^{3/2} \eta_{\mu\nu}dx^\mu dx^\nu+\left({L\over r}\right)^{3/2}{dr^2\over f(r)}
    +ds^2_5,
    \cr
    ds_5^2=~&L^{3/2}r^{1/2}\left({4r\over
    9r_0}f(r)d\theta^2
    +d\Omega_4^2\right),
    \,
    \cr
    f(r)=~&1-{r_0^3\over
    r^3}\ , \qquad\qquad L=(\pi
    Ng_s)^{1\over3}{\alpha'}^{1\over2}\ ,
    \cr
    e^\Phi=~&g_s\left(\frac rL\right)^{3/4}\ .
\end{align}
The geometry consists of a flat, warped 4d part, a radial direction $r$, a circle parameterized by $\theta$ with radius
vanishing at $r=r_0$, and a four-sphere whose volume is instead everywhere non-zero.
It is non-singular at $r=r_0$.
Notice that in the $r\to\infty$
limit the dilaton diverges: this implies  that in this limit the
completion of the present IIA model has to be found in M-theory.
The background is completed by a constant four-form field strength
\be
F_4=3L^3\omega_4\ ,
\ee
where $\omega_4$ is the volume form of the
transverse $S^4$.

An important gauge theory parameter is the KK mass scale
$1/R_{\theta}$, which is given by
\bea \label{emmezero}
\frac{1}{R_\theta}=\frac32m_0\ ,\qquad {\rm where}\qquad
m_0^2=\frac{r_0}{L^3}\ .
\eea
As can be read from the metric,
$m_0$ is also the typical glueball mass scale and, its square is proportional to the ratio between the
confining string tension $T_{QCD}$ and the UV 't Hooft coupling
$\lambda$.
As usual, the supergravity approximation is reliable  in the regime
opposite  to that in which the KK degrees of freedom can decouple from
the low energy dynamics.
The condition $T_{QCD}\ll m_0^2$ implies in fact $\lambda\ll 1$, which
is beyond the supergravity regime of validity.

%%%%%%%%%%%%%%%%%%%%%%%%%%%%%%%%%%%%%%%%%%%%%%%%%%%%%%%%%%%%%%%%%%
\section{\label{App:MSP}Mini-superspace quantization of winding strings}
In this section we elaborate on the details of the mini-superspace quantization of the WQCD and MN theories.
First we rewrite the flat subspace in \eqref{dfmn} and \eqref{defns} using cylindrical coordinates
\begin{align}
    \eta_{\mu\nu}dx^\mu dx^\nu =
    -dt^2 + dR^2 + R^2d\varphi^2 + dz^2
\end{align}
Singling out $\varphi$ as the direction the string will wind around, the mini-superspace quantization, Eq. (\ref{Eq:TheProblem}),  reduces to
\begin{itemize}
  \item MN-background
\begin{multline}
    +e^{-\frac{7}{2}\phi -2g}\partial_r\left(e^{\frac{7}{2}\phi +2g}\partial_r\Psi \right)+\partial^2_R\Psi
    +\\
    + \left(E^2-\frac{\alpha^2}{(\pi\,\alpha')^2} e^{2\phi}\,R^2\right)\Psi=0
\end{multline}
  \item WQCD-background
\begin{multline}
    +\frac{1}{L^{3}\, r^{7/4}}\partial_r\left(r^{19/4}\, f(r)\,\partial_r \Psi\right)+\partial^2_R\Psi
    +\\
    +\left(E^2-\frac{\alpha^2}{(\pi\alpha')^2} \, \frac{r^{3}}{L^{3}}\,R^2\right)\Psi=0.
\end{multline}
\end{itemize}
We can 'symmetrize' the equation using the following redefinition of the wave function
\begin{equation}
    \Psi(R,r) = \frac1{F(r)}\psi(R,r)
\end{equation}
with $F(r)=e^{\frac74\phi+g}$ and $F(r)=L^{\frac34}r^{\frac{13}{8}}f^{\frac14}(r)$ for the MN and WQCD theories respectively. This trick simplifies the Kinetic term in $r$ however now the entire equation has an overall factor of $F(r)$. This factor can be easily removed, however we now need to add a boundary condition of $\psi$, such that it vanishes at zeros of $F(r)$. In both MN and WQCD cases the wave function equation can be written in the following form :
\begin{multline}\label{App:evprob}
    \frac1{F(r)}\Bigg[-\frac{\pd^2}{\pd^2 R}-k(r)\frac{\pd}{\pd r}\left(k(r)\frac{d}{dr}\right)
    +p(r)
    +\\
    +\left(\frac{\alpha}{\alpha'\pi}\right)^2R^2q(r)
    -E^2\Bigg]\psi(R,r)=0.
\end{multline}
With
\begin{itemize}
  \item MN
  \begin{align}
    k(r)= &1
\cr
    p(r)=&~\frac74\phi''(r)+g''(r)+\left(\frac74\phi'(r)+g'(r)\right)^2
    \cr
    q(r)=&~e^{2\phi(r)}
  \end{align}
  \item WQCD
  \begin{align}
    k(r)=~&\frac1{L^{3/2}}\sqrt{r^3-r_0^3}
    \cr
    p(r)=~&L^3\Bigg(
    -\frac7{64r^2}(k(r))^2
    +\frac{7}{4r}k(r)k'(r)
    +\cr&
    +\frac14(k'(r))^2
    +\frac12k(r)k''(r)\Bigg)
    \cr
    q(r)=~&\frac{r^3}{L^3}
  \end{align}
\end{itemize}
The transformation to the canonical coordinate used in Fig. (\ref{FigPotentials}) is
\begin{equation}
    \frac{dr}{k(r)} = d\rho,
\end{equation}
Moving to the canonical coordinates is intuitive but not necessarily advantageous from a numerical point of view. In the canonical coordinates the large $r$ behavior is very steep and standard orthogonal polynomials will not give good convergence. Instead we use different set of coordinate that allow us to use Hermite polynomials:
\begin{itemize}
  \item MN
\begin{equation*}
    \frac{dr}{k(r)} =  \frac{d\rho}{\sqrt{4+\rho^2}}
\end{equation*}
  \item WQCD
\begin{equation*}
    \frac{dr}{k(r)} = \frac{d\rho}{\sqrt{\frac{\rho^4+6\rho^2r_0+12r_0^2}{8L^3}}}
\end{equation*}
\end{itemize}

%%%%%%%%%%%%%%%%%%%%%%%%%%%%%%%%%%%%%%%%%%%%%%%%%%%%%%%%%%%%%%%%%%%%%%%%%%%%%%%%%%%%%%%%
\section{\label{App:Poly}Finding the energy spectrum using Spectral methods}
%%%%%%%%%%%%%%%%%%%%%%%%%%%%%%%%%%%%%%%%%%%%%%%%%%%%%%%%%%%%%%%%%%%%%%%%%%%%%%%%%%%%%%%%%
In this appendix we describe the solution eigenvalue problems \eqref{App:evprob} using spectral methods. For a detailed description of spectral methods see \cite{Boyd01chebyshevand}. First using reparametrization we set $\phi_0=g_s N=1$ and $L=r_0=1$ in MN and WQCD, reducing both cases to a single parameter $w=\alpha/(\alpha'\pi)$. We expand the wave-functions using the wavefunctions of harmonic oscillator
\begin{align}
    \psi(R,y) =~& \sum_{n=0}^{N_{R}}\sum_{m=0}^{N_{y}}v_{(nm)}\psi_n\left(\frac{R}{l_R}\right)\pz_m\left(\frac{\rho}{l_\rho}\right),
\cr
    \psi_n(x) =~ &e^{-\frac12R^2}h_n\left(x\right),
\cr
    \pz_m(x) =~ &e^{-\frac12y^2}\hz_m\left(x\right),
\end{align}
where $h_m(x)$ are the normalized Hermite polynomial and $\hz_m\left(x\right)$ are a modified version of Hermite such that $\hz_m\left(0\right)=0$. Both sets of orthogonal polynomial can be calculated from a recursion relation:
\begin{align}
    h_{n+1}(x) =& \sqrt{\frac2{n+1}}\,xh_n(x)+\sqrt{\frac{n}{n+1}}h_{n-1}(x),
    \\
    \hz_{2n+1}(x) =&\sqrt{\frac1{n}}\,x\hz_{2n}(x)+\sqrt{\frac{2n+1}{2n}}\hz_{2n-1}(x),
    \cr
    \hz_{2n+2}(x) =&\sqrt{\frac2{2n+3}}\,x\hz_{2n+1}(x)+\sqrt{\frac{2n}{2n+3}}\hz_{2n}(x), \nonumber
\end{align}
with the initial elements
\begin{align*}
    h_0(x)=\pi^{-1/4},
    &&
    h_1(x)=\sqrt2 xh_0(x),
\cr
   \hz_{1}(x)=h_1(x),
   &&
    \hz_{2}(x)=\sqrt{\frac23}x\hz_{1}(x)
\end{align*}
Solving \eqref{App:evprob} reduces to finding eigenvalues and eigenvector of the matrix
\begin{align}
    H_{(nm)(n'm')} =~& \int_{\mathbb{R}^2}\frac{dRd\rho}{l_Rl_\rho}~
    \psi_n\left(\frac R{l_R}\right)
    \pz_m\left(\frac \rho{l_\rho}\right)
    \cdot\cr&
    \Bigg[-\frac{\pd^2}{\pd^2 R}
    -k(\rho)\frac{\pd}{\pd \rho}\left(k(\rho)\frac{\pd}{\pd \rho}\right)
    +\cr&
    +p(\rho)
    +w^2R^2q(\rho)
    \Bigg]
    \cdot\cr&
    \psi_{n'}\left(\frac R{l_R}\right)
    \pz_{m'}\left(\frac \rho{l_\rho}\right).
\end{align}
In the paper we use results for $w=1$ after checking that a order of magnitude change in $w$ give qualitatively similar result. We calculate the energy spectrum for several values of grid size $N_R=N_\rho=20,40,60,\ldots,160$. This allow us to evaluate the error of the calculation by comparing result with different grid sizes. The choice of $l_R$ and $l_\rho$ is done by optimizing for minimal errors. When the scales are properly chosen, the  errors are proportional to $N_R*N_\rho$ which matched the expectation. At large energy the wavefunction gets nearer to the large $r$ region, where the expansion is bound to fail. The reason behind that is that the equation is not separable even at large $r$ so it is impossible to get a good expansion based on a rectangular grid of 1-dimensional functions. The effect of this is that the number of reliable eigenvalues does not grow as $N_R*N_\rho$ but rather saturates (in our case) near 1500 for MN and 1000 for WQCD. Luckily the number of eigenvalues we reliably obtain is large enough for the statistical measures described in the main text.

%\bibliography{scibib}
%\bibliographystyle{Science}
\bibliography{Chaos-Hadrons}

\providecommand{\href}[2]{#2}\begingroup\raggedright\begin{thebibliography}{10}

\bibitem{0862.01040}
E.~P. Wigner, {\em {The collected works of Eugene Paul Wigner. Part A: The
  scientific papers. Volume II: Nuclear Physics. }}.
\newblock {Berlin: Springer. xi, 574 p. }, 1996.

\bibitem{Mehta:Book}
M.~L. Mehta, {\em Random Matrices}.
\newblock Elsevier, Inc., 2004.

\bibitem{Bohigas:1983er}
O.~Bohigas, M.~J. Giannoni, and C.~Schmit, {\it {Characterization of chaotic
  quantum spectra and universality of level fluctuation laws}},  {\em Phys.
  Rev. Lett.} {\bf 52} (1984) 1--4.

\bibitem{1977}
M.~V. Berry and M.~Tabor, {\it Level clustering in the regular spectrum},  {\em
  Proceedings of the Royal Society of London. Series A, Mathematical and
  Physical Sciences} {\bf 356} (1977), no.~1686 pp. 375--394.

\bibitem{PhysRevLett.48.1086}
R.~U. Haq, A.~Pandey, and O.~Bohigas, {\it Fluctuation properties of nuclear
  energy levels: Do theory and experiment agree?},  {\em Phys. Rev. Lett.} {\bf
  48} (Apr, 1982) 1086--1089.

\bibitem{Bohigas:1989rq}
O.~Bohigas and H.~A. Weidenmuller, {\it {Aspects of Chaos in Nuclear Physics}},
   {\em Ann. Rev. Nucl. Part. Sci.} {\bf 38} (1988) 421--453.

\bibitem{Pascalutsa:2002kv}
V.~Pascalutsa, {\it {A statistical analysis of hadron spectrum: Quantum chaos
  in hadrons}},  {\em Eur. Phys. J.} {\bf A16} (2003) 149--153,
  [\href{http://arxiv.org/abs/hep-ph/0201040}{{\tt hep-ph/0201040}}].

\bibitem{Markum:2005ft}
H.~Markum, W.~Plessas, R.~Pullirsch, B.~Sengl, and R.~F. Wagenbrunn, {\it
  {Quantum chaos in QCD and hadrons}},
  \href{http://arxiv.org/abs/hep-lat/0505011}{{\tt hep-lat/0505011}}.

\bibitem{Bittner:2004ff}
E.~Bittner, S.~Hands, H.~Markum, and R.~Pullirsch, {\it {Quantum chaos in
  supersymmetric QCD at finite density}},  {\em Prog. Theor. Phys. Suppl.} {\bf
  153} (2004) 295--300, [\href{http://arxiv.org/abs/hep-lat/0402015}{{\tt
  hep-lat/0402015}}].

\bibitem{Maldacena:1997re}
J.~M. Maldacena, {\it {The large N limit of superconformal field theories and
  supergravity}},  {\em Adv. Theor. Math. Phys.} {\bf 2} (1998) 231--252,
  [\href{http://arxiv.org/abs/hep-th/9711200}{{\tt hep-th/9711200}}].

\bibitem{Witten:1998qj}
E.~Witten, {\it {Anti-de Sitter space and holography}},  {\em Adv. Theor. Math.
  Phys.} {\bf 2} (1998) 253--291,
  [\href{http://arxiv.org/abs/hep-th/9802150}{{\tt hep-th/9802150}}].

\bibitem{Gubser:1998bc}
S.~S. Gubser, I.~R. Klebanov, and A.~M. Polyakov, {\it {Gauge theory
  correlators from non-critical string theory}},  {\em Phys. Lett.} {\bf B428}
  (1998) 105--114, [\href{http://arxiv.org/abs/hep-th/9802109}{{\tt
  hep-th/9802109}}].

\bibitem{'tHooft:1973jz}
G.~'t~Hooft, {\it {A Planar Diagram Theory for Strong Interactions}},  {\em
  Nucl.Phys.} {\bf B72} (1974) 461.

\bibitem{Brandhuber:1998er}
A.~Brandhuber, N.~Itzhaki, J.~Sonnenschein, and S.~Yankielowicz, {\it {Wilson
  loops, confinement, and phase transitions in large N gauge theories from
  supergravity}},  {\em JHEP} {\bf 9806} (1998) 001,
  [\href{http://arxiv.org/abs/hep-th/9803263}{{\tt hep-th/9803263}}].

\bibitem{Sakai:2004cn}
T.~Sakai and S.~Sugimoto, {\it {Low energy hadron physics in holographic QCD}},
   {\em Prog.Theor.Phys.} {\bf 113} (2005) 843--882,
  [\href{http://arxiv.org/abs/hep-th/0412141}{{\tt hep-th/0412141}}].

\bibitem{Berenstein:2002jq}
D.~E. Berenstein, J.~M. Maldacena, and H.~S. Nastase, {\it {Strings in flat
  space and pp waves from N = 4 super Yang Mills}},  {\em JHEP} {\bf 04} (2002)
  013, [\href{http://arxiv.org/abs/hep-th/0202021}{{\tt hep-th/0202021}}].

\bibitem{Gubser:2002tv}
S.~S. Gubser, I.~R. Klebanov, and A.~M. Polyakov, {\it {A semi-classical limit
  of the gauge/string correspondence}},  {\em Nucl. Phys.} {\bf B636} (2002)
  99--114, [\href{http://arxiv.org/abs/hep-th/0204051}{{\tt hep-th/0204051}}].

\bibitem{Beisert:2010jr}
N.~Beisert, C.~Ahn, L.~F. Alday, Z.~Bajnok, J.~M. Drummond, {\em et~al.}, {\it
  {Review of AdS/CFT Integrability: An Overview}},  {\em Lett.Math.Phys.} {\bf
  99} (2012) 3--32, [\href{http://arxiv.org/abs/1012.3982}{{\tt
  arXiv:1012.3982}}].

\bibitem{Gimon:2002nr}
E.~G. Gimon, L.~A. Pando~Zayas, J.~Sonnenschein, and M.~J. Strassler, {\it {A
  Soluble string theory of hadrons}},  {\em JHEP} {\bf 0305} (2003) 039,
  [\href{http://arxiv.org/abs/hep-th/0212061}{{\tt hep-th/0212061}}].

\bibitem{Bigazzi:2004ze}
F.~Bigazzi, A.~Cotrone, L.~Martucci, and L.~Pando~Zayas, {\it {Wilson loop,
  Regge trajectory and hadron masses in a Yang-Mills theory from semiclassical
  strings}},  {\em Phys.Rev.} {\bf D71} (2005) 066002,
  [\href{http://arxiv.org/abs/hep-th/0409205}{{\tt hep-th/0409205}}].

\bibitem{Hartle:1983ai}
J.~B. Hartle and S.~W. Hawking, {\it {Wave Function of the Universe}},  {\em
  Phys. Rev.} {\bf D28} (1983) 2960--2975.

\bibitem{Seiberg:1990eb}
N.~Seiberg, {\it {Notes on quantum Liouville theory and quantum gravity}},
  {\em Prog. Theor. Phys. Suppl.} {\bf 102} (1990) 319--349.

\bibitem{Douglas:2003up}
M.~R. Douglas {\em et~al.}, {\it {A new hat for the c = 1 matrix model}},
  \href{http://arxiv.org/abs/hep-th/0307195}{{\tt hep-th/0307195}}.

\bibitem{Teschner:1997fv}
J.~Teschner, {\it {The mini-superspace limit of the SL(2,C)/SU(2) WZNW model}},
   {\em Nucl. Phys.} {\bf B546} (1999) 369--389,
  [\href{http://arxiv.org/abs/hep-th/9712258}{{\tt hep-th/9712258}}].

\bibitem{Maldacena:2000hw}
J.~M. Maldacena and H.~Ooguri, {\it {Strings in AdS(3) and SL(2,R) WZW model.
  I}},  {\em J. Math. Phys.} {\bf 42} (2001) 2929--2960,
  [\href{http://arxiv.org/abs/hep-th/0001053}{{\tt hep-th/0001053}}].

\bibitem{Maldacena:2000yy}
J.~M. Maldacena and C.~Nunez, {\it {Towards the large N limit of pure N = 1
  super Yang Mills}},  {\em Phys. Rev. Lett.} {\bf 86} (2001) 588--591,
  [\href{http://arxiv.org/abs/hep-th/0008001}{{\tt hep-th/0008001}}].

\bibitem{Witten:1998zw}
E.~Witten, {\it {Anti-de Sitter space, thermal phase transition, and
  confinement in gauge theories}},  {\em Adv. Theor. Math. Phys.} {\bf 2}
  (1998) 505--532, [\href{http://arxiv.org/abs/hep-th/9803131}{{\tt
  hep-th/9803131}}].

\bibitem{springerlink:10.1007/BF01197884}
L.~A. Bunimovich, {\it On the ergodic properties of nowhere dispersing
  billiards},  {\em Communications in Mathematical Physics} {\bf 65} (1979)
  295--312. 10.1007/BF01197884.

\bibitem{Boyd01chebyshevand}
J.~P. Boyd, {\it Chebyshev and fourier spectral methods},  2001.

\bibitem{Basu:2012ae}
P.~Basu, D.~Das, A.~Ghosh, and L.~A. Pando~Zayas, {\it {Chaos around
  Holographic Regge Trajectories}},  {\em JHEP} {\bf 1205} (2012) 077,
  [\href{http://arxiv.org/abs/1201.5634}{{\tt arXiv:1201.5634}}].

\bibitem{Chamseddine:1997nm}
A.~H. Chamseddine and M.~S. Volkov, {\it {Non-Abelian BPS monopoles in N = 4
  gauged supergravity}},  {\em Phys. Rev. Lett.} {\bf 79} (1997) 3343--3346,
  [\href{http://arxiv.org/abs/hep-th/9707176}{{\tt hep-th/9707176}}].

\end{thebibliography}\endgroup
\bibliographystyle{JHEP}

 \end{document}